\documentclass[aps,prd,twocolumn,showpacs]{revtex4}

\usepackage{epsfig}
\usepackage{graphics}
\usepackage{amsmath}
\usepackage{amssymb}
\usepackage{relsize}
\usepackage{color}

\newcommand{\mathsym}[1]{{}}

\def\a{\alpha}
\newcommand{\p}[1]{\phantom{#1}}
\newcommand{\teves}{TeVeS }

\newcommand{\be}{\begin{equation}}
\newcommand{\ee}{\end{equation}}
\newcommand{\ba}{\begin{align}}
\newcommand{\ena}{\end{align}}
\newcommand{\half}{\frac{1}{2}}
\begin{document}

\title{Preferred frame parameters in the tensor-vector-scalar theory of gravity and its generalization}
\author{Eva Sagi}
\affiliation{Racah Institute of Physics, Hebrew University of
Jerusalem, Jerusalem 91904, Israel} \email{eva.sagi@mail.huji.ac.il}

\date{\today}

\begin{abstract} The Tensor-Vector-Scalar theory of gravity, which
was designed as a relativistic implementation to the modified
dynamics paradigm, has fared quite well as an alternative to dark
matter, on both galactic and cosmological scales. However, its
performance in the solar system, as embodied in the post-Newtonian
formalism, has not yet been fully investigated. Tamaki has recently
attempted to calculate the preferred frame parameters for TeVeS, but
ignored the cosmological value of the scalar field, thus concluding
that the Newtonian potential must be static in order to be
consistent with the vector equation. We show that when the
cosmological value of the scalar field is taken into account, there
is no constraint on the Newtonian potential; however, the
cosmological value of the scalar field is tightly linked to the
vector field coupling constant $K$, preventing the former from
evolving as predicted by its equation of motion. We then proceed to
investigate the post-Newtonian limit of a generalized version of
TeVeS, with {\AE}ther type vector action, and show that its
$\beta$,$\gamma$ and $\xi$ parameters are as in GR, while solar
system constraints on the preferred frame parameters $\alpha_1$ and
$\alpha_2$ can be satisfied within a modest range of small values of
the scalar and vector fields coupling parameters, and for values of
the cosmological scalar field consistent with evolution within the
framework of existing models.\end{abstract}

 \maketitle

\section{Introduction}

As is known, General Relativity (GR) cannot explain the dynamics of
our universe on large physical scales, since the amount of visible
mass clearly lies below what would be expected from the observed
gravitational effects. The usual remedy is to invoke a form of
matter which does not couple to light, therefore being referred to
as Dark Matter (DM). However, one can also take a different point of
view and modify the law of gravity itself. Such a solution to the
missing mass problem was first studied in great detail by Milgrom in
his MOND paradigm.

The modified Newtonian dynamics (MOND) paradigm~\cite{Milgrom1983},
proposes that Newtonian gravity progressively fails as accelerations
drop below a characteristic scale $\mathfrak{a}_{0}\simeq
10^{-10}\textrm{m}/\textrm{s}^2$ which is typical of galaxy
outskirts.  MOND assumes that for accelerations of order
$\mathfrak{a}_0$ or well below it, the Newtonian relation
$\mathbf{a}=-\mathbf{\nabla}\Phi_N$ is replaced by \be
\tilde{\mu}\left(|\mathbf{a}|/\mathfrak{a}_0\right)\mathbf{a}=-\mathbf{\nabla}\Phi_N,
\ee where the function $\tilde \mu(x)$ smoothly interpolates between
$\tilde\mu(x)=x$ at $x\ll 1$ and the Newtonian expectation  $\tilde
\mu(x)=1$ at $x\gg 1$. This relation with a suitable standard choice
of $\tilde\mu(x)$ in the intermediate range has proved successful
not only in justifying the asymptotical flatness of galaxy rotation
curves where acceleration scales are much below $\mathfrak{a}_0$,
but also in explaining detailed shapes of rotation curves in the
inner parts in terms of the directly seen mass, and in giving a
precise account of the observed Tully-Fisher law which correlates
luminosity of a disk galaxy with its asymptotic rotational
velocity~\cite{Bekenstein:2006}. This sharp relation, while obtained
naturally in the framework of MOND, requires quite a fine tuning of
dark halo parameters to be explained by the dark matter paradigm.

However, MOND alone is only a phenomenological prescription that
does not fulfill the usual conservation laws, nor does it make clear
if the departure from Newtonian physics is in the gravity or in the
inertia side of the equation $\mathbf{F}=m\mathbf{a}$. Moreover, it
is non relativistic, and as such it does not teach us how to handle
gravitational lensing or cosmology in the weak acceleration regimes.
To address these issues, Bekenstein designed TeVeS~\cite{BekPRD}, a
covariant field theory of gravity which has MOND as its low
velocity, weak acceleration limit, while its nonrelativistic strong
acceleration limit is Newtonian and its relativistic limit is
general relativity (GR). TeVeS sports two metrics, the ``physical''
metric on which all matter fields propagate, and the Einstein metric
which interacts with the additional fields in the theory: a timelike
dynamical vector field, $A$, and a scalar field, $\phi$. The theory
also involves a free function $\mathcal{F}$, a length scale $\ell$,
and two positive dimensionless constants $k$ and $K$. The scalar
field in TeVeS provides the additional gravitational potential for
matter, whereas the vector field provides the desired light bending
properties, in a fashion similar to the constant unit vector in
Sanders' stratified theory~\cite{SandersStratified}.

Many aspects of TeVeS have been investigated extensively, proving
the theory to be faring quite well in view of the huge challenges it
was designed to meet. Bekenstein showed that TeVeS's weak
acceleration limit reproduces MOND, and that it also has a Newtonian
limit~\cite{BekPRD}. Skordis~\cite{skordis2005} formulated the
cosmological  equations for TeVeS, for both background and linear
perturbations, and later extended his investigation to a version of
TeVeS where the action of the vector field is of Einstein-{\AE}ther
form~\cite{skordis2008}. The CMB spectrum and the matter power
spectrum $P(k)$ were calculated by Skordis, Mota, Ferreira and
Boehm~\cite{skordis:011301}, who showed how TeVeS can reproduce the
power spectrum in a manner similar to Dark Matter. Further inquiries
into TeVeS cosmology have been made by Dodelson and
Liguori~\cite{dodelson:231301}, who showed that perturbations in the
TeVeS vector field can drive structure growth, and by Bourliot et
al.,~\cite{bourliot:063508} who considered a broad family of
functions that lead to modified gravity and calculated the evolution
of the field variables both numerically and analytically.
Giannios~\cite{giannios:103511} has found exact solutions of TeVeS
for spherically symmetric systems, including Schwarzschild-like
black holes, and Sagi and Bekenstein~\cite{EvaBek} expanded upon his
work and found charged black hole solutions. Zhao and
Famaey~\cite{zhao_famaey} have put a variety of constraints on the
TeVeS free function from galaxy dynamics. Laski, Sotani and Giannios
~\cite{lasky:104019} investigated neutron stars in TeVeS, using them
to place a lower bound on the allowed cosmological value of the
scalar field, and Sotani ~\cite{sotani:064033} calculated the
fundamental oscillation modes of neutron stars for the theory,
showing how the imprint of the scalar field could be detected in
gravitational waves.

TeVeS has also been tested against a multitude of data on
gravitational lensing. Chiu, Ko and Tian have examined theoretical
predictions of TeVeS for amplifications and time delays in strong
gravitational lensing~\cite{chiu-2006-636}, while Zhao et al. have
put TeVeS predictions for image splittings and amplifications to
test against a large sample of lensed quasars~\cite{Zhao_et_al}. And
Chen and Zhao have compared the statistics of strong gravitational
lensing by galaxies with TeVeS~\cite{ChenZhao}. This work is
admirably capped by Chen, who calculated the lensing probability
with image separation larger than a given value $\Delta\theta$ in an
open, TeVeS cosmology, and showed that the predicted lensing
probabilities with the 'simple' interpolating function $x/(1+x)$
match the observational data quite well~\cite{Chen}. Angus et al.
have criticized the claim that the colliding clusters of galaxies
``the bullet'' pose a threat to gravitational lensing a la
TeVeS~\cite{angus-2006-371}.

However, it is not yet clear where the theory stands with respect to
solar system constraints, which are usually embodied in tight limits
on the allowed values of post-Newtonian (PN) parameters. Any general
metric theory of gravity can be fully characterized by ten
'parameterized post-Newtonian' (PPN) parameters~\cite{Will}, that
quantify the lowest order effects in $v^2/c^2$ and $G_NM/c^2r$. Five
of these parameters, $\zeta_1$,$\zeta_2$, $\zeta_3$, $\zeta_4$, and
$\alpha_3$, vanish identically for any 'semi-conservative' theory,
i.e. one derived, like \teves, from a covariant action principle.
Two others, known as the Eddington$-$Robertson$-$Schiff parameters
$\beta$ and $\gamma$, characterize respectively the nonlinearity and
the spatial curvature produced by gravity.  Of the remaining three
PPN parameters, two, $\alpha_1$ and $\alpha_2$ characterize
preferred frame effects, and the third, $\xi$, also known as the
Whitehead parameter, characterizes a peculiar sort of three-body
interaction. TeVeS` PPN parameters have been calculated only under
simplifying assumptions, such as spherical symmetry
~\cite{BekPRD,giannios:103511} or cosmological value of the scalar
field set to zero~\cite{Tamaki}. These assumptions did not enable
calculation of the preferred frame parameters $\alpha_1$ and
$\alpha_2$. In a theory possessing a preferred Lorentz frame, the
rest frame of the timelike vector field, one expects these
parameters to be different from zero, whereas solar system
experiments~\cite{lrr-2006-3} constrain their measured value to be
very close to zero, specifically $\alpha_1\lesssim 10^{-4}$ and
$|\a_2|\lesssim 4\times10^{-7}$. The question then rises, is there a
region of \teves  parameter space within which the preferred frame
parameters are zero as in GR, or is \teves ruled out or unreasonably
constrained by solar system tests?

In this work, we start with an overview of \teves in Section
\ref{secTeVeS} and elaborate on the PPN parameters in Section
\ref{secPPN}, giving the full set of PPN parameters for \teves with
no prior assumptions such as spherical symmetry or zero cosmological
value of the scalar field. Although we find $\beta$ and $\gamma$ to
be unity, and $\xi=0$, as in GR, we encounter a difficulty with the
vector equation, and find that to PN order, it links the
cosmological value of the scalar field to the coupling parameter of
the vector field $K$ in such a way that the scalar field is not
allowed to evolve with cosmological expansion. This is another hint
to the fact that the simple Maxwell-like action of the vector field
causes dynamical problems: it has previously been shown by
Seifert~\cite{seifert} that static spherically symmetric solutions
in this theory are unstable against spherically symmetric
perturbations, and by Contaldi et al.~\cite{Contaldi} that the
vector field runs into caustic singularities in rather generic
situations. Therefore in Section \ref{secgenTeves} we present
results for the PPN parameters for a more general form of \teves,
one with a full vector action of {\AE}ther type. We show that the
preferred frame parameters can be set to zero within a modest range
of small coupling parameters of the scalar and vector fields, and
for reasonable values of the cosmological scalar field. We conclude
that for particular ranges of the coupling parameters, \teves with
generalized vector action is indiscernible from GR in the solar
system. The full details of the calculations are given in the
appendices.

\section{TeVeS}\label{secTeVeS}

TeVeS has MOND as its weak potential, low acceleration limit, while
its weak potential, high acceleration limit is the usual Newtonian
gravity.  TeVeS is endowed with three dynamical gravitational
fields: a scalar field $\phi$, a timelike unit normalized vector
field $A^\alpha$, and the Einstein metric $g_{\alpha\beta}$ on which
the gravitational fields of the theory propagate. The theory also
employs a "physical" metric $\tilde{g}_{\alpha\beta}$ on which
gauge, spinor and Higgs fields propagate. It is related to
$g_{\alpha\beta}$ by \be\label{grelation}
\tilde{g}_{\alpha\beta}=e^{-2\phi}g_{\alpha\beta}-2A_\alpha A_\beta
\sinh(2\phi). \ee The index of $A_\alpha$ or of $\phi_{,\alpha}$ is
always raised with the metric $g^{\alpha\beta}$, the inverse of
$g_{\alpha\beta}$.

The equations of motion for the fields in TeVeS derive from a
five-term action depending on four parameters: the fundamental
gravity  constant $G$, two dimensionless parameters $k$ and $ K$ and
a fixed length scale $\ell$. The familiar Einstein-Hilbert action
for the metric and the matter action for field variables
collectively denoted $f$ have the form

\begin{eqnarray}
\label{HEaction} S_g&=&\frac{1}{16\pi G}\int
g^{\alpha\beta}R_{\alpha\beta}\, \sqrt{-g}\,d^4x,
\\
S_m&=&\int
\mathcal{L}\left(\tilde{g}_{\mu\nu},f^\alpha,f^\alpha_{;\mu},\cdot\cdot\cdot\right)\,
\sqrt{-\tilde{g}}\,d^4x. \label{matteraction}
\end{eqnarray}
Next comes the vector field's action, with $K$ a dimensionless
positive coupling constant
\begin{eqnarray}
S_v&=&-\frac{K}{32\pi G}\int
\Big[\left(g^{\alpha\beta}g^{\mu\nu}A_{[\alpha,\mu]}A_{[\beta,\nu]}\right)
\nonumber
\\
&&-\frac{2\lambda}{K}\left(g^{\mu\nu}A_\mu
A_\nu+1\right)\Big]\,\sqrt{-g}\,d^4x, \label{vectoraction}
\end{eqnarray}
which includes a constraint that forces the vector field to be
timelike (and unit normalized); $\lambda$ is the corresponding
Lagrange multiplier.  The presence of a nonzero $A^\alpha$
establishes a preferred Lorentz frame, thus breaking Lorentz
symmetry in the gravitational sector.  Finally, we have the scalar's
action ($k$ is a dimensionless positive parameter while $\ell$ is a
constant with the dimensions of length, and  ${\cal F}$ a
dimensionless free function)
 \be\label{scalaraction}S_s=-\frac{1}{2 k^2 \ell^2 G}\int
\mathcal{F}\left(k
\ell^2h^{\alpha\beta}\phi_{,\,\alpha}\phi_{,\,\beta}\right)\,\sqrt{-g}\,d^4x,
\ee Above $h^{\alpha\beta}\equiv g^{\alpha\beta}-A^\alpha A^\beta$
with $A^\alpha\equiv g^{\alpha\beta}A_\beta$.

Variation of the action with respect to $g^{\alpha\beta}$ yields the
TeVeS Einstein equations for $g_{\alpha\beta}$ \be\label{metric_eq}
G_{\alpha\beta}=8\pi G\left(
\tilde{T}_{\alpha\beta}+\left(1-e^{-4\phi}\right)A^\mu
\tilde{T}_{\mu(\alpha}A_{\beta)}+\tau_{\alpha\beta}\right)+\theta_{\alpha\beta},
\ee where $v_{(\alpha} A_{\beta)}\equiv  v_\alpha A_\beta + A_\alpha
v_\beta$, etc. The sources here are the usual matter energy-momentum
tensor $\tilde{T}_{\alpha\beta}$ (related to the variational
derivative of $S_m$ with respect to $\tilde g^{\alpha\beta}$), as
well as  the energy-momentum tensors for the scalar and vector
fields,
\begin{eqnarray}\label{tau}
\tau_{\alpha\beta}&\equiv& \frac{\mu(y)}{kG}\left(\phi_{,\,
\alpha}\phi_{, \,\beta}-A^\mu\phi_{, \mu}A_{(\alpha}\phi_{,
\,\beta)}\right)-\frac{\mathcal{F}(y)
 g_{\alpha\beta}}{2k^2 \ell^2 G}\,,
\\
\nonumber \theta_{\alpha\beta}&\equiv&
K\left(g^{\mu\nu}A_{[\mu,\,\alpha]}A_{[\nu,\,\beta]}-\frac{1}{4}g^{\sigma\tau}g^{\mu\nu}A_{[\sigma,\,\mu]}A_{[\tau,\,\nu]}g_{\alpha\beta}\right)
\\\label{theta}
&-&\lambda A_\alpha A_\beta
\end{eqnarray}
where $v_{[\alpha} A_{\beta]}\equiv  v_\alpha A_\beta - A_\alpha
v_\beta$, etc.,  and \be \mu(y)\equiv\mathcal{F}'(y);\qquad  y\equiv
kl^2 h^{\gamma\delta} \phi_{,\,\gamma}\phi_{,\,\delta}. \ee Each
choice of the function $\mathcal{F}(y)$ defines a separate TeVeS
theory. Its derivative $\mu(y)$ functions somewhat  like the
$\tilde{\mu}$ function in MOND. For $y>0$, $\mu(y)\simeq 1$
corresponds to the high acceleration, i.e., Newtonian, limit, while
the limit $0<\mu(y)\ll 1$ corresponds to the deep MOND regime.  We
shall only consider functions such that $\mathcal{F}>0$ and $\mu>0$
for either positive or negative arguments.

The equations of motion for the vector and scalar fields are
obtained by varying the  action with respect to $\phi$ and
$A_\alpha$, respectively.  We have \be \left[\mu(y)
h^{\alpha\beta}\phi_{,\,\alpha}\right]_{;\,\beta}
=kG\left[g^{\alpha\beta}+\left(1+e^{-4\phi}\right)A^\alpha
A^\beta\right] \tilde{T}_{\alpha\beta}\,, \label{scalar_eq} \ee for
the scalar and
\begin{eqnarray}
&&K A^{[\alpha;\beta]}\;_{;\beta}+\lambda A^\alpha+\frac{8\pi}{k}\mu
A^\beta\phi_{,\,\beta}g^{\alpha\gamma}\phi_{,\,\gamma}\nonumber
\\
&&=8\pi G\left(1-e^{-4\phi}\right)g^{\alpha\nu}A^\beta
\tilde{T}_{\nu\beta}\,. \label{vector_eq}
\end{eqnarray}
for the vector.   Additionally, there is the normalization condition
on the vector field \be \label{normalization} A^\alpha
A_\alpha=g_{\alpha\beta}\,A^\alpha A^\beta=-1. \ee The $\lambda$ in
Eq.~(\ref{vector_eq}), the lagrange multiplier charged with the
enforcement of the  normalization condition, can be calculated from
the vector equation.

The three parameters, $k, K$ and $\ell$, all specific to TeVeS, are
constant in the  framework of the theory, as is $G$, the fundamental
gravitational coupling constant, which does not coincide with
Newton's $G_N$.

\section{ the PPN parameters for TeVeS}\label{secPPN}

In the weak field, slow motion limit, the next-to-Newtonian order
gravitational effects of any metric gravitational theory can be
described in terms of a set of functionals of the matter variables
that answer certain criteria of "reasonableness" and simplicity (see
\cite{Will} for details), known as the Post-Newtonian potentials,
and of ten parameters, $\gamma$, $\beta$, $\zeta_1$, $\zeta_2$,
$\zeta_3$, $\zeta_4$, $\xi$, $\alpha_1$, $\alpha_2$, $\alpha_3$,
known as the PPN parameters. Since \teves has two metrics, the
Einstein metric on which the gravitational fields propagate, and the
physical matter metric, serving as background for massive and
massless matter particles, the potentials are of course to be
expressed in terms of physical coordinates defined by the physical
metric, and physical fluid variables, with indices raised and
lowered by the physical metric.

The standard form of the Post Newtonian metric is given by
\begin{align}
g_{00} =& -1+2U-2\beta
U^2-2\xi\Phi_W+\nonumber\\&+(2\gamma+2+\alpha_3+\zeta_1-2\xi)\Phi_1\nonumber \\
& +2(3\gamma-2\beta+1+\zeta_2+\xi)\Phi_2+2(1+\zeta_3)\Phi_3\nonumber\\
& +2(3\gamma+3\zeta_4-2\xi)\Phi_4-(\zeta_1-2\xi)\mathcal{A}\label{g00PPN}\\
g_{0j}= &
-\frac{1}{2}(4\gamma+3+\alpha_1-\alpha_2+\zeta_1-2\xi)V_j-\nonumber\\&-\frac{1}{2}(1+\alpha_2-\zeta_1+2\xi)W_j\label{g0jPPN}\\
g_{jk}=& (1+2\gamma U)\delta_{jk}\label{gjkPPN}
\end{align}

  The potentials are
all of the form \be
\mathfrak{U}(x,t) \equiv \int\frac{\rho({x}',t){f}({x}',t)}{|{x}-{x'}|}d^3x'\\
\ee where ${f}({x}',t)$ is given for each potential as follows
\begin{eqnarray} &&U:1,\ \ \Phi_{1}: v^{i}v^{i},\ \ \Phi_{2}: U,\ \
\Phi_{3}:\Pi,\ \ \Phi_{4}:\frac{p}{\rho},
\\
&&\Phi_{\rm W}:\int d^{3}{x''} \rho
({x''},t)\frac{({x}-{x'})_{j}}{|{x}-{x'}|^{2}} \left(
\frac{({x'}-{x''})_{j}}{|{x}-{x''}|}-\frac{\left(x-x''\right)_j}{|x'-x''|}
\right), \nonumber  \\
&&\nonumber V_{i}:v^{i}, {\cal
A}:\frac{[v_{i}({x}-{x'})_{i}]^{2}}{|{x}-{x'}|^{2}},\ \
W_{i}:\frac{v_{j}({x}-{x'})_{j}({x}-{x'})^{i}}{|{x}-{x'}|^{2}}\ .
\end{eqnarray}

The standard post Newtonian gauge is such that all dependence on
$\chi_{,00}$  and $\chi_{,ij}$ has been eliminated from $g_{00}$ and
$g_{ij}$, where $\chi$ is the "superpotential" defined as
\be\label{superpotential} \chi \equiv-\int \rho({x'},t)|{x}-{x'}|d^3
x'\ee

Thus the coordinate frame is determined up to the necessary order,
and the form of the metric components is unique.

When taking into account the cosmological value of the scalar field,
$\phi_0$, the Einstein metric and fields to post-Newtonian order
have the following form
\begin{align}
& g_{00}  =-1+h^{(1)}_{00}+h^{(2)}_{00}\label{g00exp}\\
& g_{ij}  =\left(1+h_{ij}\right)\delta_{ij}\label{gijexp}\\
& g_{0i}  =h_{0i}\label{g0iexp}\\
& A^\alpha  =(1+A^{t(1)}+A^{t(2)},A^x,A^y,A^z)\label{Aiexp}\\
& \phi  = \phi_0+\phi^{(1)}+\phi^{(2)}\label{phiexp}
\end{align}

where an index of $(1)$ denotes a term of first order in the
dimensionless gravitational potential $G_N M/c^2 r$ (and thus second
order in the velocity $v/c$), an index $(2)$ denotes a term of
second order in the potential, and $A^i$ is of $O(1.5)$ in the
potential. $\phi_0$ is to be determined by the cosmological boundary
conditions. The fluid velocity is given by \be\label{fluid_velocity}
u^\alpha=e^{-\phi_0}(1+v^{t},v^x,v^y,v^z),\ee where $v^{t}$ is
$O(1)$ and $v^i$ are $O(\frac{1}{2})$. This form of the fluid
velocity is normalized with respect to the physical metric
\be\label{fluidnorm} \tilde{g}_{\alpha\beta}u^\alpha u^\beta=-1\ee

Following the procedure described in ~\cite{Will}, one can calculate
the post-Newtonian physical metric for TeVeS. Since the calculation
is straightforward but tedious, we relegate it to the appendix. We
only mention here, that with nonzero cosmological value of the
scalar field, the physical metric is not asymptotically Minkowski,
and to be able to compare our result to the standard form, we must
transform the physical metric to local quasi-Cartesian coordinates,
through the following coordinate transformation
\be\label{coordinate_trans} x^{\bar{0}}=e^{\phi_0}x^0 \
,x^{\bar{j}}=e^{-\phi_0}x^j.\ee The potentials must then be rescaled
accordingly, for example $\bar{U}=e^{-2\phi_0}U,$
$\bar{V}_i=e^{-4\phi_0}V_i$ and $\bar{W}_i=e^{-4\phi_0}W_i.$
Additionally, we will work in units such that $G_N\equiv 1.$ In
these units, the coupling constant $G$ can be expressed in terms of
$K,k$~\cite{BekEva} \be\label{Gvalue}
G=\left(\frac{1}{1-K/2}+\frac{k}{4\pi}\right)^{-1}.\ee

 In local
quasi-Cartesian coordinates, and with $G_N\equiv 1,$ the
post-Newtonian metric for \teves is given by \be \tilde{g}_{00}=
-1+2U-2U^2+4\Phi_1+4\Phi_2+2\Phi_3+6\Phi_4,\ee\be
 \tilde{g}_{ij}=\delta_{ij}(1+2U)\ee and \be \label{g0i_solution} \tilde{g}_{0i}=
-\frac{1}{2}(7+\alpha_1-\alpha_2)V_i-\frac{1}{2}(1+\alpha_2)W_i\ee
with \be
\alpha_1=\frac{4G}{K}\left((2K-1)e^{-4\phi_0}-e^{4\phi_0}+8\right)-8\ee
and \be
\alpha_2=\frac{2G}{(2-K)^2}\left(3(2-K)-(K+4)e^{4\phi_0}\right)-1\ee

Here we omitted the bars over the potentials.
 The time-time and space-space components of the physical metric do
not depend on the cosmological value of the scalar field in the
standard post-Newtonian coordinate system, yielding
$\beta=\gamma=1$, $\xi=\zeta_i=\alpha_3=0$, with $i=1..4$, as for
GR.

The result of $\beta=1$ is apparently inconsistent with
Giannios'~\cite{giannios:103511}, who obtained $\beta\neq 1$ for
$A^r\neq 0$. However, Giannios performed the calculation assuming
spherical symmetry, and took the radial component of the vector
field to be of $O(1)$, and not $O(1.5)$ as dictated by the
Post-Newtonian formalism. While such a term is allowed by the field
equations, the vector equation to $O(1)$ is \be\nabla^2 A^i-
A^j_{,ji}=0. \ee This means that such a term in the vector field
does not originate in sources or fields, and hence it is not clear
how it would be incorporated in the PPN formalism. This was not
noticed by Giannios since he performed the calculation in vacuum, so
that the sources did not appear directly in the equations, but only
as boundary conditions for integration. An $O(1)$ term in the vector
field would require extra parameters to set the boundary conditions,
and thus seems unnatural. An $A^r$ which is $O(0.5)$ could be
acceptable, but it would have no effect on the calculation of the
PPN parameters, since it could be eliminated by an appropriate
Lorentz transformation.

However, the time-space component of the metric does depend on
$\phi_0$, as do the preferred frame parameters, and factors of
$e^{2\phi_0}$ cannot be "rescaled out" by a coordinate
transformation. Therefore it is incorrect to perform the PPN
parameters calculation without taking into account the cosmological
value of the scalar field. In fact, its role in the vector equation
is quite critical, as we shall soon see.

The covariant divergence of the vector equation, unlike the
covariant divergence of the Einstein equation, is not automatically
satisfied, but yields a constraint on the divergence of the vector
field or on the coupling constants of the theory (see ~\cite{Will},
Section 5.4 for details). In the case of \teves, we obtain to
$O(1.5)$ \be \frac{K}{1-K/2}U_{,0}=-2(1-e^{-4\phi_0})U_{,0}\ee Here
Tamaki's calculation ~\cite{Tamaki} was in error; since he assumed
$\phi_0=0$, his conclusion at this step should have been that
necessarily $K=0$, namely one cannot take into account the coupling
of the vector field in this particular theory without the scalar
field, because obviously one cannot demand $U_{,0}=0$ always. When a
nonzero cosmological value of the scalar field is taken into
account, it is constrained by the divergence of the vector equation.
From the requirement that $U_{,0}\neq 0,$ we obtain  \be
\frac{K}{1-K/2}=-2(1-e^{-4\phi_0}),\ee which links the cosmological
value of the scalar field to the coupling constant of the vector
field as follows \be\label{phi0-K-relation}
\phi_{0}=-\frac{1}{4}\ln{\left(\frac{2}{2-K}\right)}\ee

This contradicts the basic assumption of TeVeS that its coupling
parameters are constant, namely that they do not change during the
evolution of the universe. Or vice-versa, if the assumption remains
valid, then $\phi_0$ is not allowed to evolve. Moreover, this
relation would mean that $\phi_0$ is negative, since it has been
shown~\cite{EvaBek} that one must have $K<2$. This would bring the
problem of superluminal propagation of scalar waves.

When relation (\ref{phi0-K-relation}) holds, it can be shown that
there is no ambiguity in the determination of the physical metric
and the preferred frame parameters, although the spatial divergence
of the vector field remains indeterminate. However, it looks like
the constraint (\ref{phi0-K-relation}) is another sign of the
existence of a dynamical problem with the vector action in \teves,
adding up to the analysis of Contaldi et al.~\cite{Contaldi}
regarding the formation of caustic singularities in the evolution of
the vector field. Therefore, we leave aside simple \teves, and
proceed instead to calculate the PPN parameters of \teves with
{\AE}ther~\cite{AE_status} type vector action.

\section{TeVeS with {\AE}ther type vector action}\label{secgenTeves}

This version of TeVeS has been proposed by
Skordis~\cite{skordis2008}, who investigated its cosmology, as a
natural generalization, and Contaldi et al.~\cite{Contaldi} adopted
it as a resolution to the problem of vector caustic singularities.
The vector action is taken to be of the most general form quadratic
in derivatives of the vector fields, whereas its scalar and metric
actions are unaltered, thus preserving the correct MOND and
Newtonian limits. We take a vector action of the
form\begin{widetext}  \be S_v=-\frac{1}{16\pi G}\int \sqrt{-g}d^4x
\left(\frac{K}{2}F_{\alpha\beta}F^{\alpha\beta}+\frac{K_+}{2}S_{\alpha\beta}S^{\alpha\beta}+K_2\left(\nabla
A\right)^2+K_4 \dot{A}_\alpha\dot{A}^\alpha-\lambda\left(A^\alpha
A_\alpha +1\right)\right)\ee \end{widetext} where
$F_{\alpha\beta}=A_{\alpha;\beta}-A_{\beta;\alpha}$,
$S_{\alpha\beta}=A_{\alpha;\beta}+A_{\beta;\alpha}$ and
$\dot{A}^\alpha=A^\beta A^\alpha_{;\beta}$. The relation between the
coupling constants $K_i$ and the $c_i$ of {\AE}ther~\cite{AE_status}
is $c_1-c_3=2K$,$c_1+c_3=2K_+$,$c_2=K_2$ and $c_4=-K_4$.

The equation of motion and the stress tensor for the vector field
are then

\begin{widetext}
\begin{align}
&K\nabla_\alpha F^{\alpha\beta} + {K_+} \nabla_\alpha
S^{\alpha\beta} + K_2 \nabla^\beta \left( \nabla\cdot A\right) - K_4
\dot{A}^\sigma \nabla^\beta A_\sigma + K_4 \nabla_\sigma
\left(\dot{A}^\beta A^\sigma\right) + \lambda A^\beta +
\frac{8\pi}{k}\mu  A^\alpha \phi_{,\alpha} g^{\beta\gamma}
\phi_{,\gamma} \nonumber\\& =8\pi G \left(
1-e^{-4\phi}\right) g^{\beta\alpha}\tilde{T}_{\alpha\gamma} A^{\gamma}\label{new_vector}\\
&\theta_{\alpha\beta} \equiv \,K\left(F_{\sigma\alpha}F^{\sigma}_{\p{\sigma}\beta}-\frac{1}{4} F^2 g_{\alpha\beta}\right) + {K_+} \left( S_{\alpha\sigma}S_{\beta}^{\p{\beta}\sigma} - \frac{1}{4} S^2 g_{\alpha\beta} + \nabla_\sigma \left[  A^{\sigma} S_{\alpha\beta}-S^{\sigma}_{\p{\sigma} (\alpha} A_{\beta )} \right]\right)\nonumber\\
&+K_2 \left(g_{\alpha\beta} \nabla_\sigma \left( A^\sigma \nabla \cdot A \right) - A_{(\alpha}\nabla_{\beta)} \nabla \cdot A  - \frac{g_{\alpha\beta}}{2} (\nabla\cdot A)^2 \right)\nonumber\\
& + K_4 \left(\dot{A}_\beta \dot{A}_\alpha +  \dot{A}_\sigma
A_{(\alpha} \nabla_{\beta)} A^\sigma - \nabla_\sigma
\left[\dot{A}^\sigma  A_\alpha A_\beta \right]-
\frac{g_{\alpha\beta}}{2} \dot{A}_\sigma \dot{A}^\sigma \right)-
\lambda A_\alpha A_\beta\label{new_stress_tensor},
\end{align}
\end{widetext}
and the scalar equation remains unaltered. To post-Newtonian order,
the Einstein and physical metric and fields are as in Section
\ref{secPPN}, and the coordinate rescaling is as for simple TeVeS.
The new gravitational coupling constant $G_N$ is given by
\be\label{GNnew} G_N=G
\left(\left(1-\frac{K+K_+-K_4}{2}\right)^{-1}+\frac{k}{4\pi}\right),\ee
and for units in which $G_N\equiv 1$, \begin{align}
G=&\left(\left(1-\frac{K+K_+-K_4}{2}\right)^{-1}+\frac{k}{4\pi}\right)^{(-1)}\nonumber\\=&\frac{4\pi(2-(K+K_+)+K_4)}{8\pi+k(2-(K+K_+)+K_4)}\end{align}

The PPN metric is \be\tilde{g}_{00}=
-1+2U-2U^2+4\Phi_1+4\Phi_2+2\Phi_3+6\Phi_4,\ee\be
 \tilde{g}_{ij}=\delta_{ij}(1+2U)\ee and \be \tilde{g}_{0i}=
-\frac{1}{2}(7+\alpha_1-\alpha_2)V_i-\frac{1}{2}(1+\alpha_2)W_i\ee
with \begin{widetext}
\begin{align}
\alpha_1&=8\left(\frac{G\left((e^{2\phi_0}-(1-2K)e^{-2\phi_0})\sinh{\left(2\phi_0\right)}-(K_+e^{4\phi_0}+K)\right)}{2KK_+-(K+K_+)}-1\right)\\
\alpha_2&=\frac{\alpha_1}{2}+\frac{4G\left((2+3K_2+2K_+)e^{2\phi_0}-\left(2-(K+K_+-K_4)\right)e^{-2\phi_0}\right)\sinh{\left(2\phi_0\right)}}{(K_2+2K_+)\left(2-(K+K_+-K_4)\right)}
-\\&-\frac{2G\left((K_4-K+3K_+)(2+3K_2+2K_+)e^{4\phi_0}+(3K_++3K_2+K-K_4)(2-(K+K_+-K_4))\right)}{(K_2+2K_+)\left(2-(K+K_+-K_4)\right)^2}+3\end{align}\end{widetext}

Hence TeVeS with {\AE}ther type vector kinetic term has
$\beta=\gamma=1$, $\xi=\zeta_i=\alpha_3=0$, with $i=1..4$, as for
GR, and preferred frame parameters $\alpha_1$ and $\alpha_2$ given
above.

Although not apparent from the above formulation of the alphas, when
the coupling constants of TeVeS, $k$ and the $K_i$, as well as the
cosmological value of the scalar field are zero, then
$\alpha_1=\alpha_2=0$. Moreover, when the scalar field is decoupled
from the theory by setting $k=\phi_0=0$, the preferred frame
parameters acquire the values calculated for {\AE}ther theory, as
given in~\cite{foster:064015}\be
\alpha_1=\frac{4\left((K-K_+)^2-K_4(K+K_+)\right)}{2KK_+-(K+K_+)}\ee
\be
\alpha_2=\frac{\alpha_1}{2}+\frac{(K-3K_+-K_4)(K+3K_++3K_2-K_4)}{(2K_++K_2)(2-(K+K_+-K_4))}\ee
This was to be expected, since when \teves is exempted of the scalar
field, the physical and Einstein metric coincide, and the theory is
identical to {\AE}ther.

Current data strongly constrains the preferred frame parameters to
$|\alpha_1|<10^{-4}$ (from measurement of the earth-moon orbital
polarization with lunar laser ranging and from pulse timing of
binary pulsar PSR J2317+1439) and $|\alpha_2|<4\times 10^{-7}$ (from
solar alignment with the
ecliptic)~\cite{lrr-2006-3,alpha1limit,alpha2limit}. It has been
shown that in {\AE}ther theory the two alphas can be set to zero
with two parameters to spare~\cite{foster:064015}; it is obvious
that in the extended version of TeVeS, which has five parameters, we
can solve for $\alpha_1,\alpha_2=0$ with three parameters to spare.
However, since the alphas depend on the cosmological value of the
scalar field, their value is expected to change with cosmological
evolution, and the question that rises is whether the present day
cosmological value of the scalar field, as evolved from reasonable
initial conditions, is consistent with the experimentally measured
values of $\alpha_1$ and $\alpha_2$ today, and with small values of
the coupling constants $k$, $K_i$. Small values are required since
in simple \teves, in order to avoid too large MOND effects in the
solar system, $k$ should be of the order of $0.01$, and in order to
give correct lensing $K$ should also be small and of the same order
of magnitude as $k$ ~\cite{BekPRD}. It then seems natural to assume
that the extra kinetic terms added to the vector action should also
have small couplings.

There is also the question of cosmological evolution of the alphas
themselves; however, since all experimental data on the values of
the preferred frame parameters originates nearby by cosmological
standards (solar system or pulsars in our galaxy), we allow
ourselves to assume that $\phi_0$ is constant for practical
purposes.

\subsection{Allowed Ranges of Parameters}

\begin{figure}[]
\centering
\includegraphics[width=8.6cm,angle=0]{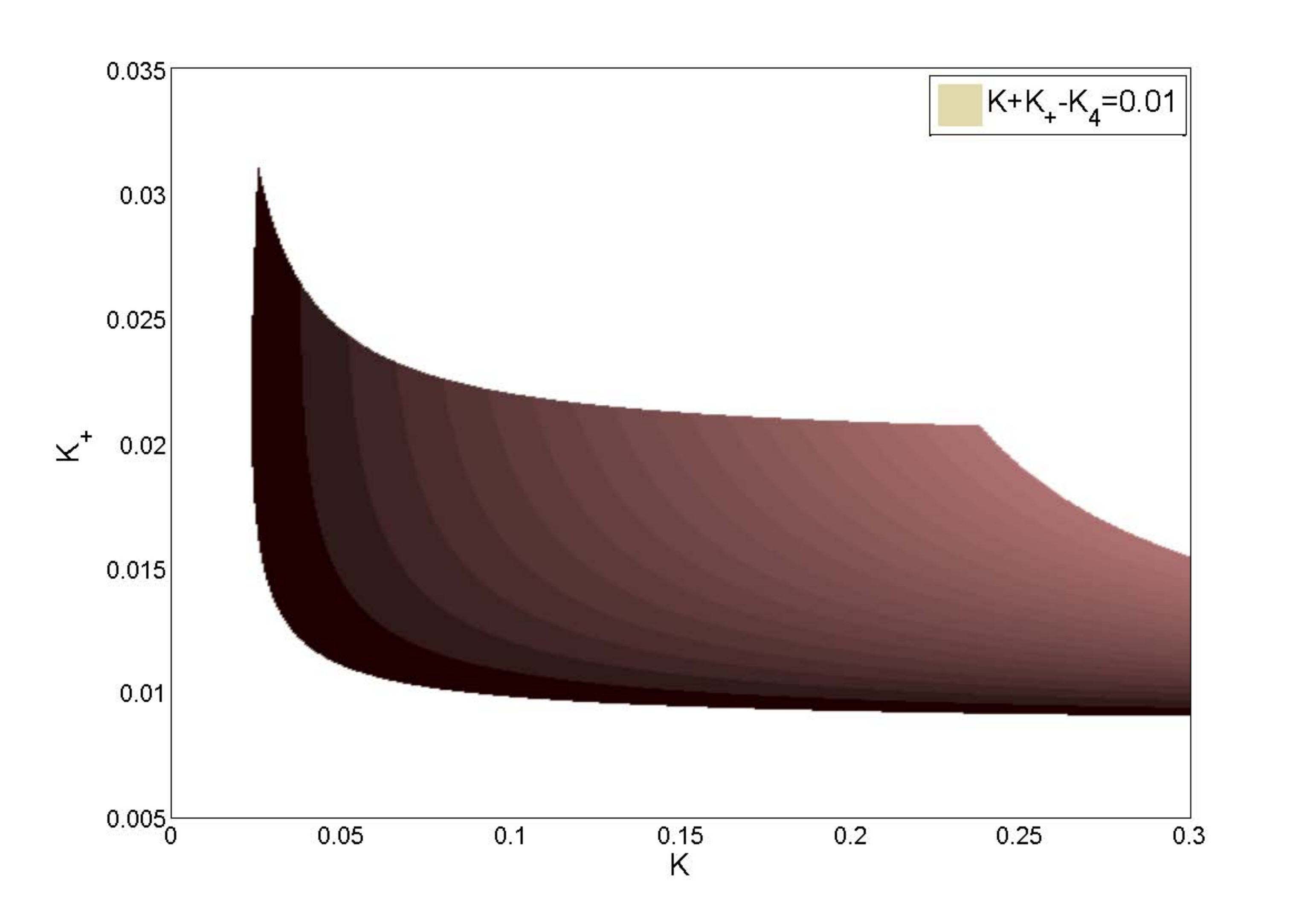}\caption{The region in the $K-K_+$ plane for which $\alpha_1,\alpha_2=0$, with $\phi_0=0.003$,
$k$ and $K_2$ in the range [0,0.3], and $K+K_+-K_4=0.01$ is shown in
color. The shading of the region shows the value of $K_2$; a lighter
hue corresponds to a higher value. $K_2$ increases with increasing
$K$, whereas $k$ increases with increasing $K_+$. \label{K2k001fig}}
\end{figure}

\begin{figure}[]
\centering
\includegraphics[width=8.6cm,angle=0]{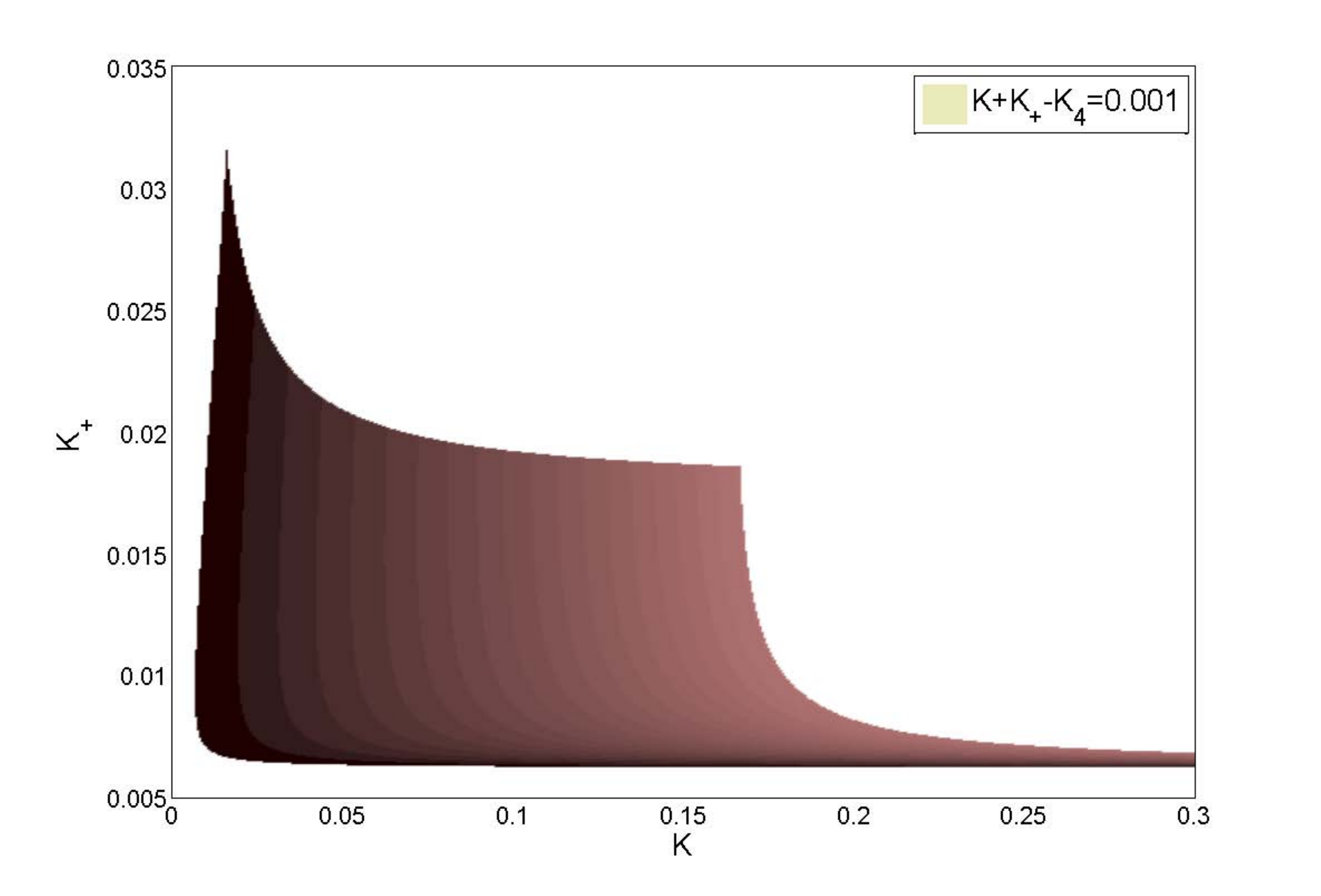}\caption{The region in the $K-K_+$ plane for which $\alpha_1,\alpha_2=0$, with $\phi_0=0.003$, $k$ and $K_2$ in the range
[0,0.3], and $K+K_+-K_4=0.001$ is shown in color. The shading of the
region shows the value of $K_2$; a lighter hue corresponds to a
higher value. $K_2$ increases with increasing $K$, whereas $k$
increases with increasing $K_+$.\label{K2k0001fig}}
\end{figure}

Solving $\alpha_1=0$ and $\alpha_2=0$, we obtained expressions for
$K_2$ and $k$ in terms of $K_4,K,K_+$ and $\phi_0$. The constraints
on the alphas are satisfied for large ranges of the parameters
$K_4,K,K_+$; however, when demanding that the parameters be small
with respect to unity, the ranges narrow significantly.  One can
look at the allowed region in the $K-K_+$ parameter space, obtained
from the overlap of the regions in which both $k$ and $K_2$ are
small, for specific values of the parameter $K_4$.  We take
$\phi_0=0.003$, as suggested by~\cite{lasky:104019}, and show, for
example, the region in the $K-K_+$ plane for which
$\alpha_1,\alpha_2=0$, $K_2$ and $k$ are in the range $[0,0.3]$,
with $K+K_+-K_4=c$, for $c=0.01$ and $c=0.001$, in
Fig.(\ref{K2k001fig}) and Fig.(\ref{K2k0001fig}), respectively.
The allowed ranges for small parameters are quite narrow, especially
for $K_+$, which is restricted to a small range near $0.01$. The
plots remain similar if instead of requiring $\alpha_1=0$ and
$\alpha_2=0$, we only demand, for instance, $\alpha_1=10^{-5}$ and
$\alpha_2=0$, in compliance with current experimental constraints.

If we increase $\phi_0$, the shape of the allowed region in the
$K-K_+$ plane remains similar, but it is shifted upwards, favoring
larger values of $K_+$. For too large a value of $\phi_0$ (around
0.01), there would be no overlap between the regions in which both
$k$ and $K_2$ are small; the same happens when $K+K_+-K_4$ is taken
to be too large. When $K+K_+-K_4=0.1$ there is no overlap between
the two regions.

In the special case $K+K_+-K_4=0$, the forms of $K_2$ and $k$ become
particularly simple\begin{widetext} \be
k=\frac{2\pi\left((2K-1)e^{-4\phi_0}-(2K_+-1)e^{4\phi_0}-2(2KK_+-(K+K_+)+1)\right)}{2KK_+-(K+K_+)}\ee
\be K_2=\frac{-2\left(K_+(2K-1)+2K\right)}{3(2K-1)}\ee\end{widetext}
Both $k$ and $K_2$ no longer depend on $K_4$, and $K_2$ doesn't
depend on $\phi_0$. However, Skordis~\cite{skordis2008} has shown
that cosmologically the combination $K+K_+-K_4$ in TeVeS with
{\AE}ther type vector action plays the same role as $K$ in simple
TeVeS, by allowing for a growing mode in the vector field in order
to source structure formation. For this same reason we have not
considered negative values of $K+K_+-K_4$.

\section{Conclusions}

We calculated the preferred frame parameters for TeVeS, and in the
process of calculation we found that the divergence of the vector
equation constrains the cosmological value of the scalar field to be
related to the coupling parameter of the vector field. Since such a
link does not allow the scalar field to evolve cosmologically, we
set aside simple TeVeS and proceeded to calculate the preferred
frame parameters for a generalized version of TeVeS previously
suggested in the literature to resolve possible dynamical
problems~\cite{skordis2008,Contaldi}, TeVeS with {\AE}ther like
vector action. We obtained expressions for $\alpha_1$ and $\alpha_2$
in terms of the coupling parameters of \teves, $k$ and $K_i$, and
the cosmological scalar field $\phi_0$. Since all existing
experimental data on values of the alphas originates from within our
galaxy, we assumed the cosmological value of the scalar field to be
constant, and of modest value, and analyzed the allowed ranges of
coupling parameters for which both preferred frame parameters are
zero. We found that the conditions $\alpha_1=0$, $\alpha_2=0$ can be
satisfied for small ranges of the coupling parameters, and in these
ranges \teves has its PPN metric identical to GR, making it
indiscernible from the latter in the solar system. Future work on
gravitational wave speeds and stability is expected to constrain the
allowed values of parameters further.

\begin{acknowledgements}
We thank Jacob D. Bekenstein for guidance and encouragement.
\end{acknowledgements}

\appendix
\section{Calculation of PPN parameters for TeVeS}\label{appPPNTeVeS}

The PPN parameters are determined by solving the gravitational field
equations with a perfect fluid source, in a standard coordinate
gauge where the spatial part of the metric is diagonal and
isotropic. The fluid variables are assigned orders of $U\sim v^2\sim
\rho\sim\Pi\sim p/\rho\sim O(1)$, where $U$ is the Newtonian
gravitational potential. Taking the time derivative of a quantity
raises its order by one half, since $\partial/\partial t\sim \nabla
v$. In the Newtonian limit of any theory of gravity, the temporal
component of the metric $g_{00}$ is required to first order in the
potential, whereas the spatial components of the metric $g_{ij}$ are
of zero order; the post-Newtonian limit calls for knowledge of
$g_{00}$ to $O(2)$, $g_{ij}$ to $O(1)$, and $g_{0j}$, which must
change sign under time reversal, is order $O(1.5)$. The standard
form of the Post-Newtonian metric is given in
Eqs.(\ref{g00PPN})-(\ref{gjkPPN}).

In the calculation we will use the following relation that holds for
$U$, $\Phi_{1,2,3,4}$, and $V_i$

\be\label{potential}
    \nabla^2 \mathfrak{U} = -4 \pi G\, \rho f.
\ee

The superpotential Eq.(\ref{superpotential}) satisfies
 \be \nabla^2 \chi=-2U\label{nabla2chieqU}\ee
  \be \chi_{,i0}=V_i-W_i\label{chiVW}\ee
  \be V_{i,i}=-U_{,0}\label{VeqU}\ee
  the last two relations follow from the continuity equation for the fluid, assumed
  to hold to   $O(1.5)$
  $$ \rho_{,0}+(\rho v^i)_{,i}=0.$$

 The standard post Newtonian gauge is such that all dependence on $\chi_{,00}$
 and $\chi_{,ij}$ has been eliminated from $g_{00}$ and $g_{ij}$.

We will follow the procedure described in Will's classic reference
~\cite{Will}. Although many details are similar to Tamaki's
computation~\cite{Tamaki}, we will show the calculation again to
make it clear where factors of $e^{\pm 2\phi_0}$ enter, and how they
alter the results. In TeVeS, which is endowed with a scalar and
vector field additional to the metric, they must be solved for to
the order necessary to derive the metric as required; namely, $\phi$
to $O(2)$ in the Newtonian gravitational potential,$A^t$ to $O(2)$
in the Newtonian gravitational potential, and $A^i$ to $O(1.5)$.
Moreover, the metric from which the PPN parameters are to be
determined is the physical metric, $\tilde{g}_{\alpha\beta}$, since
this is the metric on which matter fields propagate. For the
calculation it is convenient to rearrange the Einstein equations
(\ref{metric_eq}) as follows
\begin{eqnarray}
\label{TeVeSEqRmunu} R_{\alpha\beta}&=&\left[ 8\pi G\left(
\tilde{T}_{\mu\nu}+\left(1-e^{-4\phi}\right)A^\gamma
\tilde{T}_{\gamma (\mu}A_{\nu)}+\tau_{\mu\nu}\right)\right.
\nonumber  \\
&&+\left.\theta_{\mu\nu}
\right](\delta_{\alpha}^{\mu}\delta_{\beta}^{\nu}
-\frac{1}{2}g_{\alpha\beta}g^{\mu\nu})\ .
\end{eqnarray}
  where $\tau_{\alpha\beta}$ and $\theta_{\alpha\beta}$ are given
by Eqs. (\ref{tau})-(\ref{theta}) respectively.

For a perfect fluid source, we take as customary \be
T_{\alpha\beta}=\left(\rho+\rho\Pi+p\right)u_\alpha
u_\beta+p\tilde{g}_{\alpha\beta}\ee

In the Solar system, where accelerations are high compared to the
MOND acceleration scale $a_0$, we can take $\mu\approx 1$, and,
correspondingly, $\mathcal{F}(y)\approx y$.

With the Einstein metric and fields of the form
(\ref{g00exp})-(\ref{phiexp}), the physical metric to the required
order is given by
\begin{align} \tilde{g}_{00} &
=e^{2\phi_0}\left(-1+\left(h^{(1)}_{00}-2\phi^{(1)}\right)+\right.\nonumber\\
&\left.\left(h^{(2)}_{00}-2\phi^{(2)}-2(\phi^{(1)})^2+2\phi^{(1)}
h^{(1)}_{00}\right)\right)\label{tildegtt}\\
\tilde{g}_{ij} &
=e^{-2\phi_0}\left(1+\left(h_{ij}-2\phi^{(1)}\right)\right)\label{tildegij}\\
\tilde{g}_{0i} & =
e^{2\phi_0}h_{0i}+2A^i\sinh{2\phi_0}\label{tildeg0i}
 \end{align}
where different orders were separated by brackets.

In our calculation, we will impose the following gauge conditions
 \be
h_{kj,k}=-\frac{1}{2}\left(h^{(1)}_{00,j}-h_{kk,j}\right)\label{space_gauge}\ee
\be h_{0k,k}=\frac{1}{2-K}h_{kk,0}\label{time_gauge}\ee the first
condition is frequently used in GR, whereas the second condition is
required to bring $g_{00}$ to the standard post-Newtonian gauge.
Here and throughout the calculation we assume the Einstein summation
convention to relevant order, meaning that repeated spatial indices
are summed over. Indices on the vector field are raised and lowered
with the Einstein metric, and indices for the matter fields and
fluid velocity are raised and lowered with the physical metric.

\subsection{Calculation of $A^t$ and $v^t$ from
normalization}\label{subsecAv} We are now ready to proceed with the
calculation. First, we find the temporal components of the vector
field and of the fluid velocity from the normalization conditions
Eqs. (\ref{normalization}), (\ref{fluidnorm}), respectively \be
A^{t(1)}= \frac{1}{2}h^{(1)}_{00} \ee \be A^{t(2)}=
\frac{1}{2}h^{(2)}_{00}+\frac{3}{8}(h^{(1)}_{00})^2 \ee \be
v^{t}=\frac{1}{2}\left(h^{(1)}_{00}-2\phi^{(1)}+e^{-4\phi_0}v^2\right)\ee

\subsection{$\tilde{g}_{00}$ to $O(1)$}\label{subsecg001}

Next, we solve for $\tilde{g}_{00}$ to O(1). To this end, we must
find $h^{(1)}_{00}$ and $\phi^{(1)}$. The temporal component of the
Einstein equations (\ref{TeVeSEqRmunu}) to $O(1)$  is \be
\frac{1}{2}\left(1-K/2\right)\nabla^2 h^{(1)}_{00}= -4\pi G
e^{-2\phi_0}\rho\ee yielding for $h^{(1)}_{00}$ \be\label{h100}
h^{(1)}_{00}=\frac{2 G e^{-2\phi_0}}{1-K/2}\int
\frac{\rho}{|\mathbf{x}-\mathbf{x'}|}d^3 x'\ee For convenience, we
define \be\label{GK} G_K\equiv\frac{G}{1-K/2}\ee since this is a
recurring quantity in the calculation.

$\phi^{(1)}$ is determined from Eq.(\ref{scalar_eq}) to $O(1)$ \be
\nabla^2 \phi^{(1)}= k G e^{-2\phi_0}\rho\ee giving \be\label{phi1}
\phi^{(1)}= -\frac{k G e^{-2\phi_0}}{4\pi}\int
\frac{\rho}{|\mathbf{x}-\mathbf{x'}|}d^3 x'\ee

Combining Eqs. (\ref{h100}),(\ref{phi1}) and (\ref{tildegtt}) we
have \be \tilde{g}_{00}= G
e^{-2\phi_0}\left(\frac{1}{1-K/2}+\frac{k}{4\pi}\right)\int
\frac{\rho}{|\mathbf{x}-\mathbf{x'}|}d^3 x'\ee

We now  put the physical metric in standard Newtonian and
post-Newtonian form in local quasi-Cartesian coordinates, as
described in Ref.~\cite{Will}, through the following coordinate
transformation \be\label{coordinate_trans} x^{\bar{0}}=e^{\phi_0}x^0
\ ,x^{\bar{j}}=e^{-\phi_0}x^j\ee In the new coordinates, the
physical metric to Newtonian order is \be \tilde{g}_{00}=-1+2G_N
\bar{U} \ , \tilde{g}_{ij}=\delta_{ij} \ , \tilde{g}_{0i}=0\ee with
\be\label{GN} G_N=G \left(\frac{1}{1-K/2}+\frac{k}{4\pi}\right)\ee

Here the potential $U$ has been redefined in accordance with the
coordinate transformation $\bar{U}=e^{-2\phi_0}U.$ In the
asymptotically Minkowskian coordinates, we have
\be\label{h100solution} h^{(1)}_{00}=2 G_K \bar{U}\ee and
\be\label{phi1solution} \phi_{1}=-\frac{k G}{4\pi} \bar{U}\ee

In order to compare our result with (\ref{g00PPN}), we must select
units such that $G_N\equiv 1$. The coupling constant $G$ is then
given in terms of $K,k$ \be\label{Gvalue}
G=\left(\frac{1}{1-K/2}+\frac{k}{4\pi}\right)^{-1}\ee so that
finally \be \tilde{g}_{00}=-1+2 \bar{U} \ ,
\tilde{g}_{ij}=\delta_{ij} \ , \tilde{g}_{0i}=0\ee

\subsection{$\tilde{g}_{ij}$ to $O(1)$}\label{subsecgij}

The next step will be to solve for $\tilde{g}_{ij}$ to $O(1)$. From
the space-space component of Eq. (\ref{TeVeSEqRmunu}) we have for
$h_{ij}$ \be
\label{spatial}\nabla^2h_{ij}-h_{00,ij}+h_{kk,ij}-h_{ki,jk}-h_{kj,ik}=-8\pi
G_Ke^{-2\phi_0}{\rho}\,\delta_{ij},\ee making use of the gauge Eq.
(\ref{space_gauge}), this becomes \be \nabla^2h_{ij}=-8\pi
G_Ke^{-2\phi_0}{\rho}\,\delta_{ij},\ee whose solution is
\be\label{hij} h_{ij}=\delta_{ij}2
G_K\int\frac{{\rho}e^{-2\phi_0}}{|\mathbf{x}'-\mathbf{x}|}d^3x', \ee
or \be\label{hijsolution} h_{ij}=2{G_K}e^{-2\phi_0}{U}\delta_{ij}\ee

Combining this result with Eq. (\ref{phi1}) we obtain for
$\tilde{g}_{ij}$ \be \tilde{g}_{ij}=e^{-2\phi_0}\delta_{ij}
\left(1+2 e^{-2\phi_0} G_N{U}\right)\ee We now transform the
coordinate system according to (\ref{coordinate_trans}), set units
such that $G_N\equiv 1$, and obtain  \be \tilde{g}_{ij}=\delta_{ij}
\left(1+2 \bar{U}\right).\ee Comparing this with Eq. (\ref{gjkPPN})
we see that for TeVeS the PPN parameter $\gamma=1$. This is in
accordance with previous work~\cite{BekPRD,giannios:103511,Tamaki},
and agrees with current solar system data.

\subsection{$\tilde{g}_{0j}$ to $O(1.5)$}\label{subsecgoj}

We now solve for $\tilde{g}_{0j}$ to $O(1.5)$, which requires
calculating $A^i$ and $h_{0j}$ to the same order. First, we find the
Lagrange multiplier $\lambda$ to $O(1)$ from the temporal component
of the vector equation Eq. (\ref{vector_eq}) \be
\lambda=\frac{1}{2}K\nabla^2 h^{(1)}_{00}-16\pi G\rho
\sinh{2\phi_0}\ee

The time-space component of Eq. (\ref{TeVeSEqRmunu}) to $O(1.5)$ is
then \be\label{eqforh0j} -\frac{1}{2}\left(\nabla^2
h_{0j}-h_{0k,kj}+h_{kk,0j}-h_{kj,0k}\right)=-8\pi G e^{-6\phi_0}\rho
v^j.\ee Making use of the gauge condition Eq. (\ref{time_gauge}),
written as $h_{0k,k}=\frac{1}{2}\frac{G_K}{G}h_{kk,0}$ this becomes
\begin{align} -\frac{1}{2}\left(\nabla^2
h_{0j}-\frac{1}{2}\frac{G_K}{G}h_{kk,0j}+h_{kk,0j}-h_{kj,0k}\right)&=\nonumber\\=-8\pi
G e^{-6\phi_0}\rho v^j&.\end{align} We note that
$h_{kk}=6{G_K}e^{-2\phi_0}U$ and
$h_{kk,0j}-h_{kj,0k}=\frac{2}{3}h_{kk,0j}$, obtaining \begin{align}
-\frac{1}{2}\left(\nabla^2
h_{0j}+6{G_K}\left(\frac{2}{3}-\frac{1}{2}\frac{G_K}{G}\right)e^{-2\phi_0}U_{,0j}\right)&=\nonumber\\=-8\pi
G e^{-6\phi_0}\rho v^j.\end{align} Using the definition of the
potential $V_j$ and the relation Eq. (\ref{nabla2chieqU}), we obtain
\begin{align} -\frac{1}{2}\left(\nabla^2
h_{0j}-3{G_K}\left(\frac{2}{3}-\frac{1}{2}\frac{G_K}{G}\right)e^{-2\phi_0}\nabla^2\chi_{,0j}\right)&=\nonumber\\=2Ge^{-6\phi_0}\nabla^2
V_j.\end{align} Finally, with the aid of Eq. (\ref{chiVW}), we
rearrange and solve for $h_{0j}$ \be\label{h0jsolution}
h_{0j}=3e^{-2\phi_0}{G_K}\left(\frac{2}{3}-\frac{1}{2}\frac{G_K}{G}\right)(V_j-W_j)-4e^{-6\phi_0}{G}
V_j.\ee

Now to the spatial part of the vector. Eq. (\ref{vector_eq}) to
$O(1.5)$ is \begin{align} K\left(\nabla^2 A^i- A^j_{,ji}+\nabla^2
h_{0i}-h_{0j,ji}+\frac{1}{2}h^{(1)}_{00,i0}\right)&=\nonumber\\=-8\pi
G e^{-2\phi_0}\rho v^i(1-e^{-4\phi_0})& \end{align}

Taking the covariant divergence of the vector equation, we obtain to
$O(1.5)$

\be \frac{K}{2}h^{(1)}_{00,0ii}=-8\pi G e^{-2\phi_0}(\rho
v^i)_{,i}(1-e^{-4\phi_0}) \ee using relations
(\ref{h100solution},\ref{potential}) this can be written as \be
KG_Ke^{-2\phi_0}\nabla^2
U_{,0}=2Ge^{-2\phi_0}\nabla^2V_{i,i}(1-e^{-4\phi_0}), \ee or, with
Eq. (\ref{VeqU}) and canceling some terms on both sides, \be
\frac{K}{1-K/2}U_{,0}=-2(1-e^{-4\phi_0})U_{,0}\ee From the
requirement that $U_{,0}\neq 0,$ we obtain  \be
\frac{K}{1-K/2}=-2(1-e^{-4\phi_0}),\ee which links the cosmological
value of the scalar field to the coupling constant of the vector
field as follows \be
\phi_{0}=-\frac{1}{4}\ln{\left(\frac{2}{2-K}\right)}\ee

If the constraint Eq.(\ref{phi0-K-relation}) holds, it is possible
to solve for the vector field to $O(1.5)$ and obtain the PPN metric
and the preferred frame coefficients, and this we proceed to
demonstrate. Since the covariant divergence of the vector equation
does not constrain $A^i_{,i}$, we are free to set its value as we
please. We can regard it as a $U(1)$ gauge transformation, with
$A^\alpha\rightarrow \bar{A}^\alpha+\varphi^{,\alpha}$. Although the
action is not fully $U(1)$ gauge invariant because of the lagrange
multiplier term, to $O(1.5)$, the relevant order for the preferred
frame PPN parameters, the vector equation is gauge invariant.

We thus set, for convenience \be \label{vector_gauge} A^i_{,i}=0
,\ee simplifying the vector equation \begin{align} K\left(\nabla^2
A^i+\nabla^2
h_{0i}-h_{0j,ji}+\frac{1}{2}h^{(1)}_{00,i0}\right)&=\nonumber\\=-8\pi
G e^{-2\phi_0}\rho v^i (1-e^{-4\phi_0}) &\end{align} from Eq.
(\ref{eqforh0j}) we have \[\nabla^2 h_{0i}-h_{0j,ji}=16\pi G
e^{-6\phi_0}\rho v^i-\frac{2}{3}h_{jj,0i},\] yielding \begin{align}
K\left(\nabla^2 A^i+16\pi G e^{-6\phi_0}\rho
v^i-\frac{2}{3}h_{jj,0i}+\frac{1}{2}h^{(1)}_{00,i0}\right)&=\nonumber\\=-8\pi
G \rho v^i e^{-2\phi_0}(1-e^{-4\phi_0})&. \end{align}

Since $h^{(1)}_{00}=\frac{1}{3}h_{jj}$, we have \begin{align}
K\left(\nabla^2 A^i+16\pi G e^{-6\phi_0}\rho
v^i-\frac{1}{2}h_{jj,0i}\right)&=\nonumber\\=-8\pi G e^{-2\phi_0}
\rho v^i (1-e^{-4\phi_0})& \end{align} With the definition of the
potential $V_i$ and
\[h_{jj,0i}=-3{G_K}e^{-2\phi_0}\nabla^2
\chi_{,0i}=-3e^{-2\phi_0}{G_K}\nabla^2(V_i-W_i),\] we get
\begin{align} K \left(A^i-4e^{-6\phi_0}{G}
V_i+\frac{3}{2}e^{-2\phi_0}{G_K}(V_i-W_i)\right)&=\nonumber\\=2e^{-2\phi_0}{G}
V_i(1-e^{-4\phi_0}) \end{align}

The spatial part of the vector field is then given by \be
A^i=e^{-2\phi_0}\left(2{G}\left(2e^{-4\phi_0}+\frac{(1-e^{-4\phi_0})}{K}\right)
V_i-\frac{3}{2}{G_K}(V_i-W_i)\right),\ee or, with Eq.(\ref{GK})
\begin{align}\nonumber
A^i=2Ge^{-2\phi_0}\left(\left(2e^{-4\phi_0}+\frac{(1-e^{-4\phi_0})}{K}\right)
V_i-\right.\\\left.-\frac{3}{2-K}(V_i-W_i)\right).\label{ui_solution}\end{align}
This result satisfies $A^i_{,i}=0$ if relation
(\ref{phi0-K-relation}) holds.

 We now combine Eqs. (\ref{h0jsolution}),
(\ref{ui_solution}) and (\ref{tildeg0i}) to obtain
\begin{eqnarray}
&\tilde{g}_{0i}=2G\left(\frac{(1-2K)}{(2-K)^2}(V_i-W_i)-2e^{-4\phi_0}
V_i+\right.&\\&\nonumber\left.(1-e^{-4\phi_0})\left(\frac{1+(2K-1)e^{-4\phi_0}}{K}
V_i-\frac{3}{2-K}(V_i-W_i)\right)\right)\end{eqnarray} This remains
to be converted to asymptotically Minkowskian coordinates using
prescription (\ref{coordinate_trans}), with the potentials rescaled
accordingly $\bar{V}_i=e^{-4\phi_0}V_i$ and
$\bar{W}_i=e^{-4\phi_0}W_i,$ yielding
\begin{widetext}\be\label{g0i_solution}
\tilde{g}_{0i}=2G\left(\frac{e^{4\phi_0}(1-2K)}{(2-K)^2}(\bar{V}_i-\bar{W}_i)-2
\bar{V}_i+\left(e^{4\phi_0}-1\right)\left(\frac{1+(2K-1)e^{-4\phi_0}}{K}
\bar{V}_i-\frac{3}{2-K}(\bar{V}_i-\bar{W}_i)\right)\right)\ee\end{widetext}

In the GR limit, when $K=k=\phi_0=0$ (the limit should be taken
before the division by K in the solution to the vector equation),
and in units such that $G_N=G\equiv1$, this reduces to \be
g_{0i}=-\frac{7}{2}V_i-\frac{1}{2}W_i,\ee as expected.

Here we see again, that it is not possible to take $K=0$ without
having $\phi_0=0$, since then the physical metric would diverge.

\subsection{$\tilde{g}_{00}$ to $O(2)$}\label{subsecg002}

We have now arrived at the most complicated step- calculation of
$\tilde{g}_{00}$ to $O(2)$. To this end, we should solve for
$h^{(2)}_{00}$ from the temporal part of the Einstein equations Eqs.
(\ref{TeVeSEqRmunu}), and for $\phi^{(2)}$ from the scalar equation
Eq. (\ref{scalar_eq}) to $O(2)$. We start with $h^{(2)}_{00}$
\begin{widetext}\begin{align}\nonumber \left(1-\frac{K}{2}\right) &
\left(-\frac{1}{2}\nabla^2 h^{(2)}_{00}
+\frac{1}{2}h_{00,j}\left(h_{jk,k}-\frac{1}{2}h_{kk,j}\right)
-\frac{1}{4}|\mathbf{\nabla}h^{(1)}_{00}|^2
+\frac{1}{2}h_{jk}h^{(1)}_{00,jk}\right)\\   -\frac{1}{2}K A^j_{,j0}
&
-\frac{1}{2}\left(h_{jj,00}-2\left(1-\frac{K}{2}\right)h_{j0,j0}\right)
\nonumber  \\\label{h200_equation} & = -4\pi G
e^{-2\phi_0}\rho\left( h^{(1)}_{00}+2\phi^{(1)}\right) + 4\pi G
e^{-2\phi_0}\rho\left(\Pi+3\frac{p}{\rho}+2e^{-4\phi_0} v^2\right)
\end{align}\end{widetext}

All the terms in the second line disappear by virtue of the gauge
conditions Eqs. (\ref{time_gauge}),(\ref{vector_gauge}), leaving us
with\begin{widetext}
\begin{align}\nonumber \left(1-\frac{K}{2}\right) &
\left(-\frac{1}{2}\nabla^2 h^{(2)}_{00}
+\frac{1}{2}h_{00,j}\left(h_{jk,k}-\frac{1}{2}h_{kk,j}\right)
-\frac{1}{4}|\mathbf{\nabla}h^{(1)}_{00}|^2
+\frac{1}{2}h_{jk}h^{(1)}_{00,jk}\right)  \\ & = -4\pi G
e^{-2\phi_0}\rho\left( h^{(1)}_{00}+2\phi^{(1)}\right) + 4\pi G
e^{-2\phi_0}\rho\left(\Pi+3\frac{p}{\rho}+2 e^{-4\phi_0}v^2\right)
\end{align}\end{widetext}

The left hand side of the equation yields, similarly to GR $$
-\frac{G}{2\, G_K}\left(\nabla^2 h^{(2)}_{00}+2\left({G_K}\right)^2
e^{-4\phi_0}\nabla^2
U-8\left({G_K}\right)^2e^{-4\phi_0}\nabla^2\Phi_2\right)$$ For the
right hand side we use the definitions of the potentials
$\Phi_1$,$\Phi_2$, $\Phi_3$, $\Phi_4$, together with Eqs.
(\ref{h100solution}),(\ref{phi1solution}) to obtain
\begin{align}\nonumber&
-{G}e^{-2\phi_0}\nabla^2\left(2e^{-4\phi_0}\Phi_1\right.\\&\left.-2e^{-2\phi_0}\left({G_K}-\frac{kG}{4\pi}\right)\Phi_2+\Phi_3+3\Phi_4\right)\nonumber\end{align}

All in all, $h^{(2)}_{00}$ is given by \begin{align}\nonumber
 h^{(2)}_{00}
 = & -2e^{-4\phi_0}\left({G_K}\right)^2
U^2+4{G_K}e^{-6\phi_0}\Phi_1+\\\nonumber&\left(8\left({G_K}\right)^2-4{G_K}\left({G_K}-\frac{k}{4\pi}{G}\right)\right)e^{-4\phi_0}\Phi_2\\
& + 2{G_K}e^{-2\phi_0}\Phi_3 +
6{G_K}e^{-2\phi_0}\Phi_4\nonumber\end{align} which simplifies to
\begin{align}\nonumber h^{(2)}_{00} =&2{G_K}e^{-2\phi_0}\left(-{G_K}e^{-2\phi_0}U^2 +
2e^{-4\phi_0}\Phi_1\right.\\&\left.+2e^{-2\phi_0}G_N\Phi_2+\Phi_3+3\Phi_4\right)\end{align}

Lastly, the scalar equation to $O(2)$ is \begin{align}
&\nabla^2\phi^{(2)}-h_{jk}\phi^{(1)}_{,jk}+2k G
e^{-2\phi_0}\rho\phi^{(1)}=\nonumber\\&k G e^{-2\phi_0}\left(2\rho
e^{-4\phi_0}v^2 + \rho\Pi +3p\right)\end{align}

Using Eqs. (\ref{hijsolution}), (\ref{phi1solution}) we see that \be
h_{jk}\phi^{(1)}_{,jk}-2k G e^{-2\phi_0}\rho\phi^{(1)}=2k G
e^{-4\phi_0}\left({G_K}+\frac{k G}{4\pi} \right)U\rho,\ee and with
Eq.(\ref{GN})  this is \be h_{jk}\phi^{(1)}_{,jk}-2k G
e^{-2\phi_0}\rho\phi^{(1)}=2k G e^{-4\phi_0}G_NU\rho.\ee  We now
easily recognize the potentials $\Phi_1$,$\Phi_2$, $\Phi_3$,
$\Phi_4$ and can write \be \phi^{(2)}=-\frac{kGe^{-2\phi_0}}{4\pi
}\left(2e^{-4\phi_0}\Phi_1+2e^{-2\phi_0}G_N\Phi_2+\Phi_3+3\Phi_4\right)\ee

We now have all that is required to find $\tilde{g}_{00}$ to $O(2)$.
We first write the contribution of the $O(2)$ term \begin{align}
&h^{(2)}_{00}-2\phi^{(2)}-2(\phi^{(1)})^2+2\phi^{(1)}
h^{(1)}_{00}=2G_Ne^{-2\phi_0}\nonumber\\&\left(
2e^{-4\phi_0}\Phi_1+2e^{-2\phi_0}G_N\Phi_2+\Phi_3+3\Phi_4-e^{-2\phi_0}G_NU^2\right)\end{align}

Combining all orders of $\tilde{g}_{00}$, we get
\begin{widetext}\be \tilde{g}_{00}=
e^{2\phi_0}\left(-1+2e^{-2\phi_0}G_N\left(U-e^{-2\phi_0}G_N
U^2+2e^{-4\phi_0}\Phi_1+2e^{-2\phi_0}G_N\Phi_2+\Phi_3+3\Phi_4\right)\right).\ee\end{widetext}
Transforming coordinates according to (\ref{coordinate_trans}),
rescaling the potentials to the new coordinates and choosing units
such that $G_N=1$, we finally obtain \be \tilde{g}_{00}=
-1+2\bar{U}-2\bar{U}^2+4\bar{\Phi}_1+4\bar{\Phi}_2+2\bar{\Phi}_3+6\bar{\Phi}_4,\ee
exactly as in GR.
 From here we can instantly
read that $\alpha_{3}=\zeta_1=\zeta_2=\zeta_3=\zeta_4=0$, as
expected for equations of motion derived from a lagrangian. We also
see that $\beta=1$ and $\xi=0$.

\section{PPN parameters for TeVeS with {\AE}ther type action}

All the setup for the calculation of the PPN parameters remains as
in Section \ref{appPPNTeVeS}, except for the gauge condition for the
time-space component of the Einstein metric, which we shall
recalculate. $A^t$ and $v^t$ are as in Section \ref{subsecAv}. We
follow the same steps as in the calculation for regular \teves, and
for brevity we only show the relevant equations and altered results.
The main difference will be in the calculation of $\tilde{g}_{0i}$,
since the divergence of the vector equation will now constrain the
spatial divergence of the vector field, and not the coupling
parameters of the theory (see Ref.~\cite{Will}, Section 5.4 for
details). This will enable us to calculate $\tilde{g}_{0i}$
unambiguously, and obtain the preferred frame parameters.

\subsection{$\tilde{g}_{00}$ to $O(1)$}\label{subsecg001new}

The temporal component of the Einstein equations
(\ref{TeVeSEqRmunu}) with the altered stress tensor
Eq.(\ref{new_stress_tensor}),  is to $O(1)$\be
\frac{1}{2}\left(1-\frac{K+K_+-K_4}{2}\right)\nabla^2 h^{(1)}_{00}=
-4\pi e^{-2\phi_0} G \rho\ee yielding for $h^{(1)}_{00}$
\be\label{h100} h^{(1)}_{00}={2 e^{-2\phi_0}G
}\left(1-\frac{K+K_+-K_4}{2}\right)^{-1}\int
\frac{\rho}{|\mathbf{x}-\mathbf{x'}|}d^3 x'\ee For convenience, we
define \be\label{Gv} G_v\equiv
G\left(1-\frac{K+K_+-K_4}{2}\right)^{-1}\ee In fact, this is the
quantity that replaces $G_K$ in the calculation. With the scalar
equation unaltered, we have, as in \ref{subsecg001},
\be\label{h100solutionnew} h^{(1)}_{00}=2 G_v \bar{U}\ee and
\be\label{phi1solutionnew} \phi_{1}=-\frac{k G}{4\pi} \bar{U},\ee
with $G_v$ replacing $G_K$. The physical metric in asymptotically
Minkowski coordinates is \be \tilde{g}_{00}=-1+2G_N \bar{U} \ ,
\tilde{g}_{ij}=\delta_{ij} \ , \tilde{g}_{0i}=0\ee with Newton's
constant redefined as \be\label{GNnew} G_N=G
\left(\left(1-\frac{K+K_+-K_4}{2}\right)^{-1}+\frac{k}{4\pi}\right)\ee
\subsection{$\tilde{g}_{ij}$ to $O(1)$}\label{subsecgijnew}

The result here is again the same as in Section ~\ref{subsecgij},
with $G_v$ replacing $G_K$ \be\label{hijsolutionnew}
h_{ij}=2{G_v}e^{-2\phi_0}{U}\delta_{ij},\ee and in asymptotically
Minkowski coordinates, with $G_N\equiv 1$,  \be
\tilde{g}_{ij}=\delta_{ij} \left(1+2 \bar{U}\right).\ee Comparing
this with Eq. (\ref{gjkPPN}) we see that for TeVeS with {\AE}ther
type action, the PPN parameter $\gamma$ is still unity. This is a
promising result for the modified theory, since solar system
data~\cite{lrr-2006-3} strongly constrains the value of $\gamma$ to
be near unity.
\subsection{$\tilde{g}_{0j}$ to $O(1.5)$}\label{subsecg0jnew}

The Lagrange multiplier $\lambda$ to $O(1)$ is now given by \be
\lambda=\frac{1}{2}\left(K-K_+\right)\nabla^2 h^{(1)}_{00}-16\pi
G\rho \sinh{2\phi_0}\ee

The time-space component of Eq. (\ref{TeVeSEqRmunu}) with the new
stress tensor Eq. (\ref{new_stress_tensor}) to $O(1.5)$ is then \ba
&-\frac{1}{2}\left(\nabla^2
h_{0j}-h_{0k,kj}+h_{kk,0j}-h_{kj,0k}\right)-\frac{K_2}{2}h_{kk,0j}-\nonumber\\&-\half
\left(K_2+2K_+\right)-\left(K_2+K_+\right)A^k_{,kj}-K_+
A^j_{,kk}=\nonumber\\&-8\pi G e^{-6\phi_0}\rho
v^j\label{h0j_eq_new},\end{align} and the vector equation to
$O(1.5)$ is
\begin{align}\half
\left(K+K_+-K_4\right)h^{(1)}_{00,0j}+K\left(h_{0j,kk}-h_{0k,kj}\right)+&\nonumber\\+\half
K_2
h_{kk,0j}+K_+h_{jk,k0}+\left(K_++K_2-K\right)A^k_{,kj}+&\nonumber\\+\left(K+K_+\right)A^j_{,kk}=-8\pi
G e^{-2\phi_0}(1-e^{-4\phi_0})\rho
v^j\label{vector_eq_new}\end{align}

These two equations should be solved to yield $h_{0j}$ and $A^j$. We
still need two ingredients to be able to solve the equations
$A^j_{,j}$, which will be determined from the divergence of the
vector equation, and $h_{0k,k}$, whose value should be set from the
demand that $\tilde{g}_{00}$ be in the standard PPN gauge. In the
meanwhile, we write \be\label{metric_gauge_new}
h_{0k,k}=e^{-2\phi_0}G_v \left(3+Q\right)U_{,0},\ee and we will
solve for $Q$ in the next step of the calculation. This is inspired
by the standard GR gauge condition $h_{0k,k}=3U_{,0}$, since in the
GR limit, $G_v\rightarrow G\equiv 1$ and $\phi_0=0$. The divergence
of the vector equation is

 \ba&\half
\left(K+K_+-K_4\right)h^{(1)}_{00,0jj}+\half K_2
h_{kk,jj0}+K_+h_{jk,kj0}+\nonumber\\&+\left(K_++K_2\right)A^k_{,kjj}+K_+A^j_{,kkj}\nonumber\\&=-8\pi
G e^{-2\phi_0}(1-e^{-4\phi_0})\left(\rho v^j\right)_{,j}\end{align}
Using results from previous steps, this can be written as \ba
-\half(3K_++K+3K_2-K_4)e^{-2\phi_0}G_v\chi_{,0jjkk}&\nonumber\\+(2K_++K_2)A^j_{,jkk}=2Ge^{-2\phi_0}(1-e^{-4\phi_0})V_{j,jkk}\end{align}
which simplifies to \ba
A^j_{,j}=\frac{Ge^{-2\phi_0}}{2K_++K_2}\left(2(1-e^{-4\phi_0})V_{j,j}\right.&\nonumber\\\left.+\frac{3K_++K+3K_2-K_4}{2-(K+K_+-K_4)}\chi_{,0jj}\right)\end{align}
and finally, with $V_{j,j}=-U_{,0}=\half \chi_{,0jj}$, we have for
the spatial divergence of the vector field \be\label{vec_div}
A^j_{,j}=\frac{Ge^{-2\phi_0}}{2K_++K_2}\left(1-e^{-4\phi_0}+\frac{3K_++K+3K_2-K_4}{2-(K+K_+-K_4)}\right)\chi_{,0jj}.\ee
The addition of extra kinetic terms in the vector action has removed
the $U(1)$ gauge invariance, and enabled us to determine
unambiguously the spatial divergence of the vector field.

We may now substitute the vector divergence Eq.(\ref{vec_div}) and
the metric gauge Eq.(\ref{metric_gauge_new}) into Eqs.
(\ref{vector_eq_new},\ref{h0j_eq_new}) and solve for $A^i$ and
$h_{0i}$. Doing so, we
obtain\begin{widetext}\begin{align}h_{0i}=&\frac{4Ge^{-2\phi_0}(Ke^{-4\phi_0}+K_+)V_i}{2KK_+-(K+K_+)}-\nonumber\\&\frac{Ge^{-2\phi_0}\left(2K(2-(K+K_+-K_4))e^{-4\phi_0}+
(2KK_+-(K+K_+))Q+K_+-3K+2K_+(K_4-K_++2K)\right)\chi_{,0i}}{(2-(K+K_+-K_4))(2KK_+-(K+K_+))}\\
\nonumber
A^i=&\frac{2Ge^{-2\phi_0}((1-2K)e^{-4\phi_0}-1)}{2KK_+-(K+K_+)}V_i+
Ge^{-2\phi_0}\left(\frac{(K+(2K-1)(K_++K_2))e^{-4\phi_0}}{(K_2+2K_+)(2KK_+-(K+K_+))}+\right.\\
&\left.\frac{4(1-K)K_+^2-2(K_4+K_2(3K-2)+1)K_++2(1+2K_2)K-(K_4+2)K_2}{(K_2+2K_+)(2KK_+-(K+K_+))(2-(K+K_+-K_4))}\right)\chi_{,0i}\end{align}
\end{widetext}

We still need to find $Q$; to this end, we proceed to the last step
in the calculation.

\subsection{$\tilde{g}_{00}$ to $O(2)$}\label{subsecg002new}

The equation for $h_{00}^{(2)}$ is
\begin{widetext}\begin{align}\nonumber
&\left(1-\frac{K+K_+-K_4}{2}\right)\left(-\frac{1}{2}\nabla^2
h^{(2)}_{00}
+\frac{1}{2}h_{00,j}\left(h_{jk,k}-\frac{1}{2}h_{kk,j}\right)
-\frac{1}{4}|\mathbf{\nabla}h^{(1)}_{00}|^2
+\frac{1}{2}h_{jk}h^{(1)}_{00,jk}\right)\\&
-\frac{1}{2}\left(K+3K_2+3K_+-K_4\right) A^j_{,j0}
-\frac{1}{2}\left(1+\frac{3}{2}K_2+K_+\right)h_{jj,00}+\half\left(2-(K+K_+-K_4)\right)h_{j0,j0}\
\nonumber  \\ & = -4\pi e^{-2\phi_0} G\rho\left(
h^{(1)}_{00}+2\phi_{(1)}\right) + 4\pi e^{-2\phi_0} G
\rho\left(\Pi+3\frac{p}{\rho}+2 e^{-4\phi_0}v^2\right)
\end{align}\end{widetext} All terms on the second line are proportional to
$\chi_{,00}$; therefore the standard PPN gauge is obtained from the
requirement \begin{widetext}\be
-\frac{1}{2}\left(K+3K_2+3K_+-K_4\right) A^j_{,j}
-\frac{1}{2}\left(1+\frac{3}{2}K_2+K_+\right)h_{jj,0}+\half\left(2-(K+K_+-K_4)\right)h_{j0,j}=0\ee\end{widetext}
or \begin{widetext}\be h_{j0,j}=
\frac{\left(1+\frac{3}{2}K_2+K_+\right)}{2-(K+K_+-K_4)}h_{jj,0}+
+\frac{Ge^{-2\phi_0}\left(3K_++K+3K_2-K_4\right)}{2K_++K_2}\left(1-e^{-4\phi_0}+\frac{3K_++K+3K_2-K_4}{2-(K+K_+-K_4)}\right)\chi_{,0jj}\ee\end{widetext}
Combining this with (\ref{metric_gauge_new}), we obtain for $Q$\be
Q=\frac{3K_++K+3K_2-K_4}{2K_++K_2}\left(e^{-4\phi_0}+\frac{2-4K_+}{2-(K_++K-K_4)}\right)\ee

Using this value of $Q$, in quasi-Cartesian coordinates, with the
potentials appropriately rescaled, the time-space component of the
physical metric is given by
\begin{widetext}\begin{align}\nonumber\tilde{g}_{0i}=&\left(\frac{2\left((2+3K_2+2K_+)e^{2\phi_0}-\left(2-(K+K_+-K_4)\right)e^{-2\phi_0}\right)\sinh{\left(2\phi_0\right)}}{(K_2+2K_+)\left(2-(K+K_+-K_4)\right)}\right.\\
&\nonumber-\frac{2\left((e^{2\phi_0}-(1-2K)e^{-2\phi_0})\sinh{\left(2\phi_0\right)}-(K_+e^{4\phi_0}+K)\right)}{2KK_+-(K+K_+)}\\
&\nonumber\left.-\frac{\left((K_4-K+3K_+)(2+3K_2+2K_+)e^{4\phi_0}+(3K_++3K_2+K-K_4)(2-(K+K_+-K_4))\right)}{(K_2+2K_+)\left(2-(K+K_+-K_4)\right)^2}\right)GV_i+\\
+&\nonumber\left(\frac{\left((K_4-K+3K_+)(2+3K_2+2K_+)e^{4\phi_0}+(3K_++3K_2+K-K_4)(2-(K+K_+-K_4))\right)}{(K_2+2K_+)\left(2-(K+K_+-K_4)\right)^2}\right.\\
&\nonumber-\frac{2\left((2+3K_2+2K_+)e^{2\phi_0}-\left(2-(K+K_+-K_4)\right)e^{-2\phi_0}\right)\sinh{\left(2\phi_0\right)}}{(K_2+2K_+)\left(2-(K+K_+-K_4)\right)}\\
&\left.-\frac{2\left((e^{2\phi_0}-(1-2K)e^{-2\phi_0})\sinh{\left(2\phi_0\right)}-(K_+e^{4\phi_0}+K)\right)}{2KK_+-(K+K_+)}\right)G
W_i  \label{g0j_solution_new}\end{align}
\end{widetext}\


In the standard PPN gauge, for TeVeS with {\AE}ther type vector
kinetic term, the equation for $h^{(2)}_{00}$ is the same as for
simple TeVeS, Eq.(\ref{h200_equation}), except for the replacement
of $G_K$ by $G_v$. Since the scalar equation remains unchanged, we
can immediately write the result for $\tilde{g}_{00}$, after
coordinate redefinition and appropriate rescaling of the potentials,
and this time with
$G_N=G\left(\left(1-\frac{K+K_+-K_4}{2}\right)^{-1}+\frac{k}{4\pi}\right)\equiv1$
\be \tilde{g}_{00}=
-1+2\bar{U}-2\bar{U}^2+4\bar{\Phi}_1+4\bar{\Phi}_2+2\bar{\Phi}_3+6\bar{\Phi}_4,\ee
once again we have $\alpha_{3}=\zeta_1=\zeta_2=\zeta_3=\zeta_4=0$,
as expected for equations of motion derived from a lagrangian, and
also $\beta=1$ and $\xi=0$, as for simple TeVeS. We can now compare
Eq.(\ref{g0j_solution_new}) to the standard form Eq. (\ref{g0jPPN}),
and extract the preferred frame parameters\begin{widetext}
\begin{align}
\alpha_1&=8\left(\frac{G\left((e^{2\phi_0}-(1-2K)e^{-2\phi_0})\sinh{\left(2\phi_0\right)}-(K_+e^{4\phi_0}+K)\right)}{2KK_+-(K+K_+)}-1\right)\\
\alpha_2&=\frac{\alpha_1}{2}+\frac{4G\left((2+3K_2+2K_+)e^{2\phi_0}-\left(2-(K+K_+-K_4)\right)e^{-2\phi_0}\right)\sinh{\left(2\phi_0\right)}}{(K_2+2K_+)\left(2-(K+K_+-K_4)\right)}
-\\&-\frac{2G\left((K_4-K+3K_+)(2+3K_2+2K_+)e^{4\phi_0}+(3K_++3K_2+K-K_4)(2-(K+K_+-K_4))\right)}{(K_2+2K_+)\left(2-(K+K_+-K_4)\right)^2}+3\end{align}\end{widetext}
In the above,
$$G=\frac{4\pi(2-(K+K_+)+K_4)}{8\pi+k(2-(K+K_+)+K_4)}.$$


\end{document}